\documentclass {revtex4}
\usepackage[english]{babel} %
\usepackage {hyperref}
\usepackage {amsmath,amssymb,caption,fancyhdr,graphicx,lineno,remreset,sidecap,url}
\usepackage{amssymb}
\usepackage{amsmath}
\usepackage{multirow}

\usepackage{minitoc}

\begin{document}
\title{HEISENBERG UNCERTAINTY PRINCIPLE AND ECONOMIC ANALOGUES OF BASIC PHYSICAL QUANTITIES}

\author{Soloviev V., Prof. Dr. Sc.}
\email{vnsoloviev@rambler.ru}
\affiliation{Cherkasy National University named after B. Khmelnitsky, Cherkassy, Ukraine}

\author{Saptsin V., PhD.}
\email{saptsin@sat.poltava.ua}
\affiliation{Kremenchug National University named after M. Ostrogradskii, Kremenchuk, Ukraine}


\begin{abstract}
From positions, attained by modern theoretical physics in understanding of the universe bases, the methodological and philosophical analysis of fundamental physical concepts and their formal and informal connections with the real economic measurings is carried out. Procedures for heterogeneous economic time determination, normalized economic coordinates and economic mass are offered, based on the analysis of time series, the concept of economic Plank's constant has been proposed. The theory has been approved on the real economic dynamic's time series, including stock indices, Forex and spot prices, the achieved results are open for discussion.
\end{abstract}




\maketitle
\label{contents}
\tableofcontents

\section{Introduction}
\label{sec:Intro}

The instability of global financial systems depending on ordinary and natural disturbances in modern markets and highly undesirable financial crises are the evidence of methodologial crisis in modelling, predicting and interpretation of current socio-economic conditions.

In papers \cite{r001SolSapSimferopolThesis,r002SaptsinSoloviev,r003SapSolArxiv} we have suggestsed a new paradigm of complex systems modelling based on the ideas of quantum as well as relativistic mechanics. It has been revealed that the use of quantum-mechanical analogies (such as the uncertainty principle, notion of the operator, and quantum measurement interpretation) can be applied to describing socio-economic processes.
In papers \cite{r001SolSapSimferopolThesis,r002SaptsinSoloviev,r003SapSolArxiv} we have suggestsed a new paradigm of complex systems modelling based on the ideas of quantum as well as relativistic mechanics. It has been revealed that the use of quantum-mechanical analogies (such as the uncertainty principle, notion of the operator, and quantum measurement interpretation) can be applied to describing socio-economic processes.

It is worth noting that quantum analogies in economy need to be considered as the subject of new inter-disciplinary direction -- quantum econophysics (e.g. \cite{r004BqaqueQuantumFin2004,r005MaslovQuantumEconomics06,r006GuevaraQuantumEconophysics06,r007,r008Goncalez2011}), which, despite being relatively young, has already become a part of classical econophysics \cite{r009MantegnaStanley00, r010Romanovsky07, r011solovievmatheconomics, r012SolDerbMonogr2010}.

Ideas \cite{r001SolSapSimferopolThesis,r002SaptsinSoloviev,r003SapSolArxiv} were anticipated and further developed in our works on modelling, predicting, and identification of socio-economic systems \cite{r013SaptsinGenComplMarkovChains08, r014SaptsinMarkovPaper09, r015SolSapChDrezden09, r016ChabM09, r017ChabM10, r018ChabRiga10, r019SolSapChPsepGlava, r020KuznetzKoleb2011} (complex Markov chains), \cite{r019SolSapChPsepGlava, r020KuznetzKoleb2011, r021Fourier2009, r022SapChFourierKharkov, r023ChabS10} (discrete Fourier Transorm), \cite{r024SapDvufaznye05, r025OlhovayaKonkur2010, r026SapChabAGranUstoych11, r027SapSolBatyrNelnConcur2agents10} (multi-agent modelling), \cite{r028SolSapDynNetwMaths, r029SolSapChabAPedZbornik} (dynamic network mathematics), \cite{r030SaptsinSetPrir, r031SaptsinMerezhPrirody} (network measurements), \cite{r032SolSapShokotko, r033HeisenbergRiga} (uncertainty principle in economics) etc.

However significant differences between physical and socio-economical phenomena, diversity and complexity of mathematical toolset (which, on account of historical circumstances, has been developed as the language of sciences), as well as lack of deep understanding of quantum ideology among the scientists, working at the joint of different fields require a special approach and attention while using quantum econophysical analogies.

\textbf{Aim of work. }Our aim is to conduct detailed methodological and phylosophical analysis of fundamental physical notions and constants, such as time, space and spatial coordinates, mass, Planck's constant, light velocity from the point of view of modern theoretical physics, and search of adequate and useful analogues in socio-economic phenomena and processes.

\section{About nature and interrelations of basic physical notions}
\label{sec:NatBasicPhysNotions}

Time, distance and mass are normally considered to be initial, main or basic physical notions, that are not strictly defined. It is thought that they can be matched with certain numerical values. In this case other physical values, e.g. speed, acceleration, pulse, force, energy, electrical charge, current etc. can be conveyed and defined with the help of the three above-listed ones via appropriate physical laws.

Let us emphasize that none of the modern physical theories, including relativistic and quantum physics, can exist without basic notions. Nevertheless, we would like to draw attention to the following aspects.

As Einstein has shown in his relativity theory, presence of heterogeneous masses leads to the distortion of 4-dimensional time-space in which our world exists. As a result Cartesian coordinates of the 4-dimensional Minkowski space $(x,y,z,ict)$, including three ordinary Cartesian coordinates $(x,y,z)$ and the forth formally introduced time-coordinate $ict$ ($i=\sqrt{-1} $ - imaginary unit, $c$ - speed of light in vacuum, $t$ - time), become curvilinear \cite{r034landau1975classical}.

It is also possible to approach the interpretation of Einstein's theory from other point of view, considering that the observed heterogeneous mass distribution is the consequence of really existing curvilinear coordinates $(x,y,z,ict)$. Then the existence of masses in our world becomes the consequence of geometrical factors (presence of time-space and its curvature) and can be described in geomatrical terms.

If we step away from global macro-phenomena that are described by the general relativity theory, and move to micro-world, where laws of quantum physics operate, we come to the same conclusion about the priority of time-space coordinates in the definition of all other physical values, mass included.

To demonstrate it, let us use the known Heisenberg's uncertainty ratio which is the fundamental consequence of non-relativistic quantum mechanics axioms and appears to be (e.g. \cite{r002SaptsinSoloviev}):
\begin{equation}
\label{eq:GrindEQ__1_}
 \Delta x\cdot \Delta v\ge \frac{\hbar }{2m_{0} } , 
\end{equation}
where $\Delta x$ and $\Delta v$ are mean square deviations of  $x$ coordinate and velocity $v$ corresponding to the particle with (rest) mass $m_{0} $, $\hbar $ - Planck's constant. Considering values $\Delta x$ и $\Delta v$ to be measurable when their product reaches its minimum, we derive (from (\ref{eq:GrindEQ__1_})):
\begin{equation}
\label{eq:GrindEQ__2_}
 m_{0} =\frac{\hbar }{2\cdot \Delta x\cdot \Delta v} , 
\end{equation}
i.e. mass of the particle is conveyed via uncertainties of its coordinate and velocity -- time derivative of the same coordinate.

Nowadays, scientists from different fields occupy themselves with the investigation of structure and other fundamental properties of spacetime from physical, methodological, psychological, philosophical and other points of view. However, theoretical physics \cite{r035CramerTransactional86, r036kaku1999introduction}, including its most advanced and developing spheres (e.g. string theory \cite{r036kaku1999introduction, r037BalasubramanianWhatWeDontKnow}) is expected to show the most significant progress in understanding the subject, though there is no single concept so far \cite{r035CramerTransactional86, r036kaku1999introduction, r037BalasubramanianWhatWeDontKnow, r038Vladimirov_p1, r039Vladimirov_p2, r040kaku2006parallel, r041kaku2008physics}.

Within fundamental physical science we can mark out two investigational directions: 1) receipt of quantitive patterns, possible to verify experimentally or empirically and 2) interpretation of existing theories or development of new theories, that allow accurate and laconic (involving as little as possible mathematical notions and formalisms) interpretation of basic physical facts. The second direction is especially important when speaking of transferring physical notions and mathematical formalisms into other spheres, e.g. economics.

Not claiming to be exhaustive, aiming to make the audience (professional economists included) as wide as possible, we will confine ourselves to the examination of the most typical and clear examples.

According to the concept \cite{r038Vladimirov_p1, r039Vladimirov_p2}, having been developed for the last couple of decades by the Moscow school of theoretical physicists (headed by Y. Vladimirov), space, time, and four fundamental physical interactions (gravitational, electromagnetic, strong and weak) are secondary notions. They share common origins and are generated by the so-called world matrix which has special structure and peculiar symmetrical properties. Its elements are complex numbers which have double transitions in some abstract pre-space.

At the same time, physical properties of spacetime in this very point are defined by the nonlocal (``immediate'') interaction of this point with its close and distant neighbourhood, and acquire statistical nature. In other words, according to Vladimirov's concept, the observed space coordinates and time have statistical nature.

It is worth noting that similar ideas as of interpreting quantum mechanics, different from those of the Copenhagen school were proclaimed by John Cramer \cite{r035CramerTransactional86} (Transactional interpretation of quantum mechanics).

In our opinion the afore-metioned conception of nonlocal statistical origin of time and space coordinates can be qualitatively illustrated on the assuptions of quantum-mechanical uncertainty principle using known ratios (e.g. \cite{r002SaptsinSoloviev}:)
\begin{equation}
\label{eq:GrindEQ__3_}
 \Delta p\cdot \Delta x\sim \hbar ; 
\end{equation}

\begin{equation}
\label{eq:GrindEQ__4_}
 \Delta E\cdot \Delta t\sim \hbar ; 
\end{equation}

\begin{equation}
\label{eq:GrindEQ__5_}
 \Delta p\cdot \Delta t\sim \frac{\hbar }{c} .  
\end{equation}
 Interpreting values $\Delta E,\Delta p,\Delta x,\Delta t$ as uncertainties of particle's energy $E$, its pulse $p$,  coordinate $x$ and time localization $t$ (the latter ratio relates to the relativistic case $E=pc$, and is formally derived from the ratio (\ref{eq:GrindEQ__4_}), if $\Delta E=\Delta p\cdot c$, and takes into account maximum speed $c$ limitations in an explicit form), let us conduct the following reasoning.

 While $\Delta x\to 0$ uncertainty of pulse, and thus particle energy, uncertainty, formally becomes as big as possible, which can be provided only by its significant and nonlocal energetical interaction with the rest of the neighbourhood \ref{eq:GrindEQ__3_}. On the other side, while  $\Delta p\to 0$ the particle gets smeared along the whole space (according to (\ref{eq:GrindEQ__3_}) $\Delta x\to \infty $), i.e. becomes delocalized. It might be supposed that the fact of ``delocalized'' state of the particle takes place in any other, not necessarily marginal  $\Delta x$ and $\Delta p$ value ratios.

Similar results can be acquired while analyzing ratios (\ref{eq:GrindEQ__4_})-(\ref{eq:GrindEQ__5_}), and for temporary localization $\Delta t$.

 Vladimirov's concept probably becomes more graphic (at least for those, who are familiar with the basics of the band theory), if one remembers that so-called ``electrones'' and ``holes'' are considered to be really existing charge bearers in semiconductors. These ``particles'' have negative and positive charge respectively, accurate to the decimal place, which corresponds to the charge of a free electron, and are characterised by effective masses $m_{e} $ and $m_{h} $, different from the mass of a free electron (generally $m_{e} $ and $m_{h} $ can also be tensor values). However, in reality, these particles are virtual results of the whole semicondoctor crystal -- so-called quasi-particles -- and don't exist beyond its bounds.

Drawing the analogy with crystal it can be supposed that all structural formations of our Universes are such ``quasi-particles'', caused by nonlocal interaction and non-existent beyond its spacetime bounds. 

In conclusion we would like to note that the conept of nonlocal interaction is quite capable of giving the logical explanation to the empirical fact of indistinguishability and identity of all microparticles of this kind, which always takes place during the observation (identification) regardless of spacetime localization of this very observation.

\section{Dynamical peculiarities of economic measurements, economical analog of Heisenberg's uncertainty ratio}
\label{sec:DynEconMeasurements}

Main physical laws are normally distinguished with the presence of constants, that have been staying unchanged for the past  $\sim 10^{11} $ years (the age of our Universe since so-called ``big bang''- the most widespread hypothesis of its origin). Gravitational constant, speed of light in vaccuum, Planck's constant are among the above-listed.

Speaking of economic laws, based on the results of both physical (e.g. quantities of material resources) and economical (e.g. their value) dynamic measurements, the situation will appear to be somewhat different. Adequacy of the formalisms used for mathematical descriptions has to be constantly checked and corrected if necessary. The reason is that measurements always imply a comparison with something, considered to be a model, while there are no constant standards in economics (they change not only quantitavely, but also qualitatively -- new standards and models appear). Thus, economic measurements are fundamentally relative, are local in time, space and other socio-economic coordinates, and can be carried out via consequent and/or parallel comparisons ``here and now'', ``here and there'', ``yesterday and today'', ``a year ago and now'' etc. (see \cite{r030SaptsinSetPrir, r031SaptsinMerezhPrirody} for further information on the subject). 

Due to these reasons constant monitoring, analysis, and time series prediction (time series imply data derived from the dynamics of stock indices, exchange rates, spot prices and other socio-economic indicators) becomes relevant for evaluation of the state, tendencies, and perspectives of global, regional, and national economies. 

Let us proceed to the description of structural elements of our work and building of the model.

Suppose there is a set of $M$ time series, each of $N$ samples, that correspond to the single distance $T$, with an equal minimal time step $\Delta t_{\min } $:
\begin{equation}
\label{eq:GrindEQ__6_}
 X_{i} (t_{n} ),\begin{array}{c} {} \end{array}t_{n} =\Delta t_{\min } n;\begin{array}{c} {} \end{array}n=0,1,2,...N-1;\begin{array}{c} {} \end{array}i=1,2,...M. 
\end{equation}
To bring all series to the unified and non-dimentional representation, accurate to the additive constant, we normalize them, having taken a natural logarithm of each term of the series:
\begin{equation}
\label{eq:GrindEQ__7_}
 x_{i} (t_{n} )=\ln X_{i} (t_{n} ),\begin{array}{c} {} \end{array}t_{n} =\Delta t_{\min } n;\begin{array}{c} {} \end{array}n=0,1,2,...N-1;\begin{array}{c} {} \end{array}i=1,2,...M. 
\end{equation}
Let us consider that every new series $x_{i} (t_{n} )$ is a one-dimensional trajectory of a certain fictitious or abstract particle numbered $i$, while its coordinate is registered after every time span $\Delta t_{\min } $, and evaluate mean square deviations of its coordinate and speed in some time window $\Delta T$:
\begin{equation}
\label{eq:GrindEQ__8_}
 \Delta T=\Delta N\cdot \Delta t_{\min } =\Delta N,\begin{array}{c} {} \end{array}1<<\Delta N<<N.
\end{equation}
 The ``immediate'' speed of $i$ particle at the moment $t_{n} $ is defined by the ratio:
\begin{equation}
\label{eq:GrindEQ__9_}
 v_{i} \left(t_{n} \right)=\frac{x_{i} (t_{n+1} )-x_{i} (t_{n} )}{\Delta t_{\min } } =\frac{1}{\Delta t_{\min } } \ln \frac{X_{i} (t_{n+1} )}{X_{i} (t_{n} )} ,  
\end{equation}
its variance $D_{v_{i} } $:
\begin{equation}
\label{eq:GrindEQ__10_}
 D_{v_{i} } =\frac{1}{(\Delta t_{\min } )^{2} } \left(<\ln ^{2} \frac{X_{i} (t_{n+1} )}{X_{i} (t_{n} )} >_{n,\Delta N} -\left(<\ln \frac{X_{i} (t_{n+1} )}{X_{i} (t_{n} )} >_{n,\Delta N} \right)^{2} \right), 
\end{equation}
and mean square deviation $\Delta v_{i} $:
\begin{equation}
\label{eq:GrindEQ__11_}
 \Delta v_{i} =\sqrt{D_{v_{i} } } =\frac{1}{(\Delta t_{\min } )} \left(<\ln ^{2} \frac{X_{i} (t_{n+1} )}{X_{i} (t_{n} )} >_{n,\Delta N} -\left(<\ln \frac{X_{i} (t_{n+1} )}{X_{i} (t_{n} )} >_{n,\Delta N} \right)^{2} \right)^{\frac{1}{2} } , 
\end{equation}
where $<...>_{n,\Delta N} $ means averaging on the time window of $\Delta T=\Delta N\cdot \Delta t_{\min } $ length. Calculated according to (\ref{eq:GrindEQ__11_}) value of $\Delta v_{i} $ has to be ascribed to the time, corresponding with the middle of the avaraging interval $\Delta T$.

 To evaluate dispersion $D_{x_{i} } $ coordinates of the $i$ particle are used in an approximated ratio:
\begin{equation}
\label{eq:GrindEQ__12_}
 2D_{x_{i} } \approx D_{\Delta x_{i} } , 
\end{equation}
where 
\[D_{\Delta x_{i} } =\begin{array}{c} {} \end{array}<\left(x_{i} (t_{n+1} )-x_{i} (t_{n} )\right)^{2} >_{n,\Delta N} -\left(<x_{i} (t_{n+1} )-x_{i} (t_{n} )>_{n,\Delta N} \right)^{2} =\] 
\begin{equation}
\label{eq:GrindEQ__13_}
  =\begin{array}{c} {} \end{array}<\ln ^{2} \frac{X_{i} (t_{n+1} )}{X_{i} (t_{n} )} >_{n,\Delta N} -\left(<\ln \frac{X_{i} (t_{n+1} )}{X_{i} (t_{n} )} >_{n,\Delta N} \right)^{2} , 
\end{equation}
which is derived from the supposition that $x$ coordinates neighbouring subject to the time of deviation from the average value $\bar{x}$ are weakly correlated:
\begin{equation}
\label{eq:GrindEQ__14_}
 <\left(x_{i} \left(t_{n} \right)-\bar{x}\right)\left(x_{i+1} \left(t_{n} \right)-\bar{x}\right)>_{n,\Delta N} \approx 0. 
\end{equation}
Thus, taking into account \ref{eq:GrindEQ__12_} and \ref{eq:GrindEQ__13_} we get:
\begin{equation}
\label{eq:GrindEQ__15_}
 \Delta x_{i} =\sqrt{\frac{D_{\Delta x_{i} } }{2} } =\frac{1}{\sqrt{2} } \left(<\ln ^{2} \frac{X_{i} (t_{n+1} )}{X_{i} (t_{n} )} >_{n,\Delta N} -\left(<\ln \frac{X_{i} (t_{n+1} )}{X_{i} (t_{n} )} >_{n,\Delta N} \right)^{2} \right)^{\frac{1}{2} } .
\end{equation}
 Pay attention that it was not necessary for us to prove the connection \ref{eq:GrindEQ__12_}, as it was possible to postulate statement (\ref{eq:GrindEQ__15_}) as the definition of $\Delta x_{i} $. 

 It is also worth noting that the value

\[\left|v_{i} \left(t_{n} \right)\right|\cdot \Delta t_{\min } =\left|\ln \frac{X_{i} (t_{n+1} )}{X_{i} (t_{n} )} \right|,\] 

which, accurate to multiplier $\Delta t_{\min } $ coincides with $\left|v_{i} \left(t_{n} \right)\right|$ (see (\ref{eq:GrindEQ__9_})), is commonly named absolute returns, while dispersion of a random value $\ln \left({X_{i} (t_{n+1} )\mathord{\left/ {\vphantom {X_{i} (t_{n+1} ) X_{i} (t_{n} )}} \right. \kern-\nulldelimiterspace} X_{i} (t_{n} )} \right)$, which differs from $D_{v_{i} } $ by $(\Delta t_{\min } )^{2} $ (see (\ref{eq:GrindEQ__13_})) -- volatility.

 The chaotic nature of real time series allows to $x_{i} (t_{n} )$ as the trajectory of a certain abstract quantum particle (observed at $\Delta t_{\min } $ time spans). Analogous to (\ref{eq:GrindEQ__1_}) we can write an uncertainty ratio for this trajectory:
\begin{equation}
\label{eq:GrindEQ__16_}
\Delta x_{i} \cdot \Delta v_{i} \sim \frac{h}{m_{i} } , 
\end{equation}
or, taking into account (\ref{eq:GrindEQ__11_}) and (\ref{eq:GrindEQ__15_}): 
\begin{equation}
\label{eq:GrindEQ__17_}
 \frac{1}{\Delta t_{\min } } \left(<\ln ^{2} \frac{X_{i} (t_{n+1} )}{X_{i} (t_{n} )} >_{n,\Delta N} -\left(<\ln \frac{X_{i} (t_{n+1} )}{X_{i} (t_{n} )} >_{n,\Delta N} \right)^{2} \right)\sim \frac{h}{m_{i} } ,  
\end{equation}
where $m_{i} $ - economic ``mass'' of an $i$ series, $h$ - value which comes as an economic Planck's constant.

 Having rewritten the ration \ref{eq:GrindEQ__17_}:

\begin{equation} 
\label{eq:GrindEQ__18_} 
\Delta t_{\min } \cdot \frac{m_{i} }{(\Delta t_{\min } )^{2} } \left(<\ln ^{2} \frac{X_{i} (t_{n+1} )}{X_{i} (t_{n} )} >_{n,\Delta N} -\left(<\ln \frac{X_{i} (t_{n+1} )}{X_{i} (t_{n} )} >_{n,\Delta N} \right)^{2} \right)\sim h 
\end{equation} 

and interpreting the multiplier by $\Delta t_{\min } $ in the left part as the uncertainty of an ``economical'' energy (accurate to the constant multiplier), we get an economic analog of the ratio (\ref{eq:GrindEQ__4_}).

Since the analogy with physical particle trajectory is merely formal, $h$ value, unlike the physical Planck's constant $\hbar $, can, generally speaking, depend on the historical period of time, for which the series are taken, and the length of the averaging interval (e.g. economical processes are different in the time of crisis and recession), on the series number $i$ etc. Whether this analogy is correct or not depends on particular series' roperties.

Let us generalize the ratios (\ref{eq:GrindEQ__17_}), (\ref{eq:GrindEQ__18_}) for the case, when economic measurements on the time span $T$, used to derive the series (\ref{eq:GrindEQ__6_}), are conducted with the time step $\Delta t=k\cdot \Delta t_{\min } $, where $k\ge 1$ - is a certain given integer positive number. From the formal point of view it would mean that all terms, apart from those numbered $n=0,k,2k,3k,...$. are discarded from the initial series (\ref{eq:GrindEQ__6_}). As a result the ratios would be the following: 
\begin{equation} 
\label{eq:GrindEQ__19_} 
 \frac{1}{k\Delta t_{\min } } \left(<\ln ^{2} \frac{X_{i} (t_{n+k} )}{X_{i} (t_{n} )} >_{n,\Delta N} -\left(<\ln \frac{X_{i} (t_{n+k} )}{X_{i} (t_{n} )} >_{n,\Delta N} \right)^{2} \right)\sim \frac{h}{m_{i} } , 
\end{equation} 
\begin{equation}
 \label{eq:GrindEQ__20_} 
 k\Delta t_{\min } \cdot \frac{1}{(k\Delta t_{\min } )^{2} } \left(<\ln ^{2} \frac{X_{i} (t_{n+k} )}{X_{i} (t_{n} )} >_{n,\Delta N} -\left(<\ln \frac{X_{i} (t_{n+k} )}{X_{i} (t_{n} )} >_{n,\Delta N} \right)^{2} \right)\sim \frac{h}{m_{i} } 
\end{equation} 

and would be dependent on $k$.

 Let us proceed to the analysis of the acquired results, that have to be considered as intermediate.

 In case of $h=const$, the formal analogy with the physical particle would be complete, and in this case, as appears from (\ref{eq:GrindEQ__19_}), variance  of a random $i$-numbered value:

\[\ln \frac{X_{i} (t_{n+k} )}{X_{i} (t_{n} )} \approx \frac{X_{i} (t_{n+k} )-X_{i} (t_{n} )}{X_{i} (t_{n} )} \] 

- practically coinciding with the relative increment of terms of the $i$ initial series -- would keep increasing in a linear way with $k\Delta t_{\min } $ (interval between the observations) growing. Such dynamics is peculiar to the series with statistically independent increments.

However, both in cases of a real physical particle and its formal economic analogue any kind of change influences on the result. Therefore statistic properties of the ``thinned'' series, used to create the ratio (\ref{eq:GrindEQ__19_}), have to depend on \textit{real }measurements in the intermediate points if there were any. Besides, presence of ``long'' and ``heavy'' ``tails'' increasing along the amplitude with decreasing $\Delta t$ on distributions of corresponding returns ${\Delta X\mathord{\left/ {\vphantom {\Delta X X}} \right. \kern-\nulldelimiterspace} X} $, are in our opinion the evidence of this thesis (see for example \cite{r031SaptsinMerezhPrirody}).

 Thus, generalizing everything said above, ${h\mathord{\left/ {\vphantom {h m_{i} }} \right. \kern-\nulldelimiterspace} m_{i} } $ratio on the right side of (\ref{eq:GrindEQ__19_}) (or (\ref{eq:GrindEQ__20_})) has to be considered a certain unknown function of the series number $i$, size of the averaging window$\Delta N$, time $\bar{n}$ (centre of the averaging window), and time step of the observation (registration) $k$.

To get at least an approximate, yet obvious, formula of this function and track the nature of dependencies, we postulate the following model presentation of the right side (\ref{eq:GrindEQ__19_}):
\begin{equation}
 \label{eq:GrindEQ__21_} 
 \frac{h}{m_{i} } \simeq \frac{\tau \left(\bar{n},\Delta N_{\tau } \right)\cdot H_{i} \left(k,\bar{n},\Delta N_{H} \right)}{\Delta t_{\min } \cdot m_{i} } , 
\end{equation} 
where
\begin{equation}
 \label{eq:GrindEQ__22_} 
 \frac{1}{m_{i} } =\begin{array}{c} {} \end{array}<\varphi _{i} \left(n,1\right)>_{(0\le n\le N-2)} ,
\end{equation} 
$m_{i} $ is a non-dimentional  economic mass of an $i$-numbered series,

\begin{equation} \label{eq:GrindEQ__23_} \tau \left(\bar{n}\right)=\frac{<\varphi _{i} \left(n,1,\Delta N_{\tau } \right)>_{(\bar{n}-{\Delta N_{\tau } \mathord{\left/ {\vphantom {\Delta N_{\tau }  2\begin{array}{c} {} \end{array}<n<\begin{array}{c} {} \end{array}\bar{n}+{\Delta N_{\tau } \mathord{\left/ {\vphantom {\Delta N_{\tau }  2}} \right. \kern-\nulldelimiterspace} 2} }} \right. \kern-\nulldelimiterspace} 2\begin{array}{c} {} \end{array}<n<\begin{array}{c} {} \end{array}\bar{n}+{\Delta N_{\tau } \mathord{\left/ {\vphantom {\Delta N_{\tau }  2}} \right. \kern-\nulldelimiterspace} 2} } ),\begin{array}{c} {} \end{array}(1\le i\le M)} }{<\left(<\varphi _{i} \left(n,1,\Delta N_{\tau } \right)>_{(\bar{n}-{\Delta N_{\tau } \mathord{\left/ {\vphantom {\Delta N_{\tau }  2\begin{array}{c} {} \end{array}<n<\begin{array}{c} {} \end{array}\bar{n}+{\Delta N_{\tau } \mathord{\left/ {\vphantom {\Delta N_{\tau }  2}} \right. \kern-\nulldelimiterspace} 2} }} \right. \kern-\nulldelimiterspace} 2\begin{array}{c} {} \end{array}<n<\begin{array}{c} {} \end{array}\bar{n}+{\Delta N_{\tau } \mathord{\left/ {\vphantom {\Delta N_{\tau }  2}} \right. \kern-\nulldelimiterspace} 2} } ),\begin{array}{c} {} \end{array}(1\le i\le M)} \right)>_{\bar{n}} }  
\end{equation} 

- local physical time compression ($\tau \left(\bar{n}\right)<1$) or magnification ($\tau \left(\bar{n}\right)>1$) ratio, which allows to introduce the notion of heterogenous economic time (for a homogenous $\tau \left(\bar{n}\right)=1$),

\begin{equation}
 \label{eq:GrindEQ__24_} H_{i} \left(k,\bar{n}\right)=\frac{<\varphi _{i} \left(n,k,\Delta N_{H} \right)>_{\bar{n}-{\Delta N_{H} \mathord{\left/ {\vphantom {\Delta N_{H}  2\begin{array}{c} {} \end{array}<n<\begin{array}{c} {} \end{array}\bar{n}+{\Delta N_{H} \mathord{\left/ {\vphantom {\Delta N_{H}  2}} \right. \kern-\nulldelimiterspace} 2} }} \right. \kern-\nulldelimiterspace} 2\begin{array}{c} {} \end{array}<n<\begin{array}{c} {} \end{array}\bar{n}+{\Delta N_{H} \mathord{\left/ {\vphantom {\Delta N_{H}  2}} \right. \kern-\nulldelimiterspace} 2} } } }{<\varphi _{i} \left(n,1,\Delta N_{H} \right)>_{\bar{n}-{\Delta N_{H} \mathord{\left/ {\vphantom {\Delta N_{H}  2\begin{array}{c} {} \end{array}<n<\begin{array}{c} {} \end{array}\bar{n}+{\Delta N_{H} \mathord{\left/ {\vphantom {\Delta N_{H}  2}} \right. \kern-\nulldelimiterspace} 2} }} \right. \kern-\nulldelimiterspace} 2\begin{array}{c} {} \end{array}<n<\begin{array}{c} {} \end{array}\bar{n}+{\Delta N_{H} \mathord{\left/ {\vphantom {\Delta N_{H}  2}} \right. \kern-\nulldelimiterspace} 2} } } } ;\begin{array}{c} {} \end{array}k=1,2,...k_{\max } 
 \end{equation} 

- non-dimentional coefficient of the order of unit, which indicates differences in the dependence of variance $D_{\Delta x_{i} } $ (see (\ref{eq:GrindEQ__13_}) taking into account the case of $k\ge 1$) on the law $D_{\Delta x_{i} } \sim k$ for the given $i$ and $\bar{n}$.

\begin{equation} 
\label{eq:GrindEQ__25_} 
\varphi _{i} \left(n,k,\tilde{N}\right)=\frac{1}{k} \left(\ln ^{2} \frac{X_{i} (t_{n+k} )}{X_{i} (t_{n} )} -\left(<\ln \frac{X_{i} (t_{n+k} )}{X_{i} (t_{n} )} >_{n,\tilde{N}} \right)^{2} \right)
\end{equation} 

(index $\tilde{N}=N,\Delta N_{\tau } ,\Delta N_{H} $ in the last formula indicates the averaging parameters  according to $n$ and formulae (\ref{eq:GrindEQ__22_}),(\ref{eq:GrindEQ__23_}),(\ref{eq:GrindEQ__24_}), averaging windows $\Delta N_{\tau } ,\Delta N_{H} $ are chosen with thew following the conditions taken into consideration:
\begin{equation} 
\label{eq:GrindEQ__26_} 
 k_{\max } <\Delta N_{\tau } <\Delta N_{H} <N. 
\end{equation} 

 According to the definitions (\ref{eq:GrindEQ__23_}),(\ref{eq:GrindEQ__24_}) for coefficients $\tau \left(\bar{n}\right)$ and $H_{i} \left(k,\bar{n}\right)$ following conditions of the normalization take place:
\begin{equation} 
\label{eq:GrindEQ__27_} 
 <\tau \left(\bar{n}\right)>_{\bar{n},N} =1;\begin{array}{c} {} \end{array}H_{i} \left(1,\bar{n}\right)=1,
\end{equation} 

and the multiplier ${1\mathord{\left/ {\vphantom {1 \Delta t_{\min } }} \right. \kern-\nulldelimiterspace} \Delta t_{\min } } $ on the right side (\ref{eq:GrindEQ__21_}) can be considered as an invariant component of an economic Planck's constant $h$:
\begin{equation} 
\label{eq:GrindEQ__28_} 
 \bar{h}={1\mathord{\left/ {\vphantom {1 \Delta t_{\min } }} \right. \kern-\nulldelimiterspace} \Delta t_{\min } } ,
\end{equation} 

As you can see, $\bar{h}$ has a natural dimension <time> to the negative first power.

 It is also worth noting that average economic mass of the whole set of series (or any separate group of the series) can be introduced with the help of the following formula: 
\begin{equation} 
\label{eq:GrindEQ__29_} 
 \frac{1}{m} =\frac{1}{M} \sum _{i=1}^{M}\frac{1}{m_{i} }  .
\end{equation} 

 Acquired with the help of series (\ref{eq:GrindEQ__6_}) ratios (\ref{eq:GrindEQ__7_},(\ref{eq:GrindEQ__19_})-
 (\ref{eq:GrindEQ__28_}) also allow different interpretations. For example, it can be considered that normalized series \eqref{eq:GrindEQ__7_} depict the trajectory of a certain hypothetical economic quantum quasi-particle in an abstract $M$-dimensional space of economic indices, and $m_{i}^{-1} $ are the main components of  inverse mass tensor of the quasi-particle (the analogy with quasi-particles as free carriers of the electric charge in semiconductors), which has already been used in the previous chapter.

 In the final part of this chapter we would like to pay attention to the chosen variant of the theory, which is probably the simplest one, because of the following reasons. 

 Carrying out various $n$ (discrete time)  and $i$ (series number) averagings, we didn't take into account at least two fairly important factors: 1) amounts of financial and material resources (their movement is reflected by each series) and 2) possible correlation between the series. 

 However generalization of the theory and introduced notions is not so difficult in this case. It is enough to form a row $\left(\alpha _{1} ,\alpha _{2} ,...\alpha _{M} \right)$ of positive weight coefficients with the following condition of normalization:
\begin{equation} 
\label{eq:GrindEQ__30_} 
 \sum _{i=1}^{M}\alpha _{i}  =M,
\end{equation} 
with each of them taking into account the importance of separate series in terms of a certain criterion, while for random values

\begin{equation} 
\label{eq:GrindEQ__31_} 
\phi _{i} \left(n,k\right)=\sqrt{\frac{\alpha _{i} }{k} } \ln \frac{X_{i} (t_{n+k} )}{X_{i} (t_{n} )} ,\begin{array}{c} {} \end{array}i=1,2,...M 
\end{equation} 

instead of a one-dimensional massive (\ref{eq:GrindEQ__25_}) we should introduce a covariance matrix:
\begin{equation} 
\label{eq:GrindEQ__32_} 
 \Psi =\left[\psi _{ij} \right], 
\end{equation} 
where
\begin{equation} 
\label{eq:GrindEQ__33_} 
 \psi _{ij} =\psi _{ij} \left(k,\tilde{N}\right)=\begin{array}{c} {} \end{array}<\left(\phi _{i} \left(n,k\right)-\bar{\phi }_{i} \left(n,k\right)\right)\cdot \left(\phi _{j} \left(n,k\right)-\bar{\phi }_{j} \left(n,k\right)\right)>_{n,\tilde{N}} , 
\end{equation} 

\begin{equation} 
\label{GrindEQ__34_} 
\bar{\phi }_{i} \left(j,k,\tilde{N}\right)=\begin{array}{c} {} \end{array}<\phi _{i} \left(n,k\right)>_{n,\tilde{N}}  
\end{equation} 

(with $\alpha _{i} =1$ and absence of correlations $\psi _{ij} =\varphi _{i} \delta _{ij} $). Using a standard algorithm of characteristic constants  $\lambda _{i} ,\begin{array}{c} {} \end{array}i=1,2,...M$ and corresponding orthonormal vectors $C_{i} =\left(c_{i1} ,c_{i2} ,...c_{iM} \right)$ search in $\Psi $ matrix, we proceed to the new basis, where ``renormalized'' series $y_{i} (t_{n} )=\sum _{j=1}^{M}c_{ij} x_{j} \left(t_{n} \right) $ (new basis vectors) aren't correlated any more. However the presence of zero characteristic constants or $\lambda _{i} $, which are distinguished with relatively low values in absolute magnitude, will mean that the real dimension of the set of series (\ref{eq:GrindEQ__7_}) is in fact less than $M$ (initial series (\ref{eq:GrindEQ__7_}) or their parts are strongly correlated). In this case renormalized series $y_{i} (t_{n} )$ with zero or low characteristic constants have to be discarded. The remaining renormalized series will undergo all above-listed procedures.

\section{Experimental results and their discussion}
\label{sec:Experiments}

To test the suggested ratios and definitions we have chosen 9 economic series with $\Delta t_{\min } $ in one day for the period from April 27, 1993 to March 31, 2010. The chosen series correspond to the following groups that differ in their origin:

1) stock market indices: USA (S\&P500), Great Britain (FTSE 100) and Brazil (BVSP);

2) currency dollar cross-rates (chf, jpy, gbp);

3) commodity market (gold, silver, and oil prices).

On Fig. 1-3 normalized plots of the corresponding series, divided by groups, are introduced, while $\Delta t_{\min } $is taken equal to the unit.

As you can see from the Fig. 1-3, all time series include visually noticeable chaotic component and obviously differ from each other, which allows us to hope for the successful application the afore-mentioned theory to the interpretation and analysis of real series. Let us confine to its elementary variant.

As an example on fig. 4 we suggest absolute values of immediate speeds  (or absolute returns according to the general terminology used in literature), calculated with the help of the formula evaluation (\ref{eq:GrindEQ__9_}), and their variance (volatility), calculated with the help of the formula evaluation (\ref{eq:GrindEQ__13_}) for the series of Japanese yen (jpy) US dollar cross-rates.

 As we can see from the plots, the dependence of immediate speed or returns on time is of chaotic nature, while the dependence of volatility is smooth but not monotonous. For the rest of initial series, the dependencies of volatility and returns are similar to the depicted on the fig. 4 ones.

\begin{figure}[H]
\centerline{\includegraphics[width=6.16in,height=4.35in]{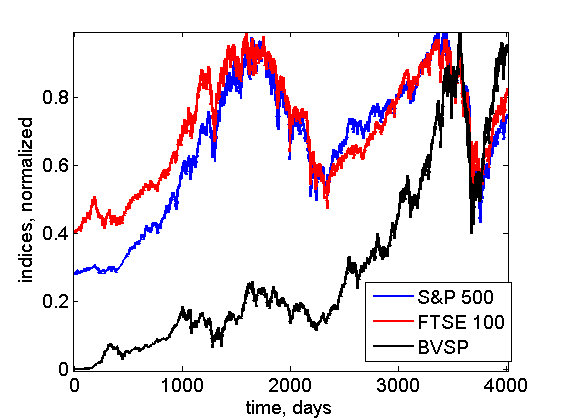}}
\caption{USA (S\&P500), Great Britain (FTSE 100), and Brazil (BVSP) daily stock indices from April 27, 1993 to March 31, 2010.}
\label{fig:Fig01}
\end{figure}

\begin{figure}[H]
\centerline{\includegraphics[width=6.16in,height=4.35in]{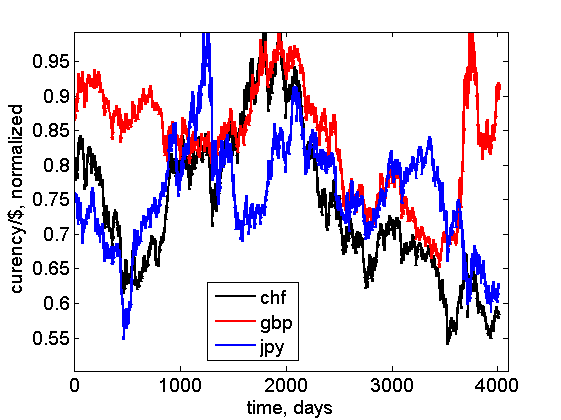}}
\caption{Daily currency dollar cross-rates (chf, jpy, gbp)}
\label{fig:Fig02}
\end{figure}

\begin{figure}[H]
\centerline{\includegraphics[width=6.16in,height=4.35in]{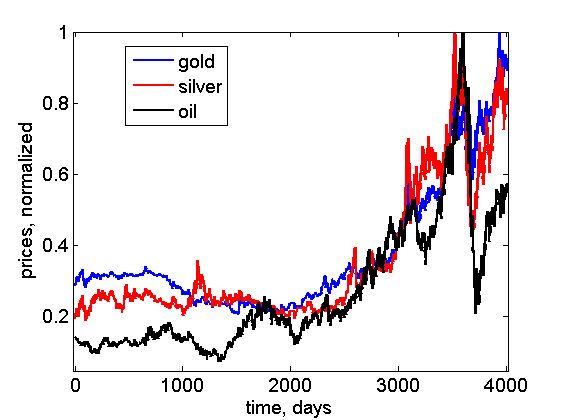}}
\caption{Commodity market. Daily gold, silver, and oil prices.}
\label{fig:Fig03}
\end{figure}

\begin{figure}[H]
\centerline{\includegraphics[width=6.16in,height=4.35in]{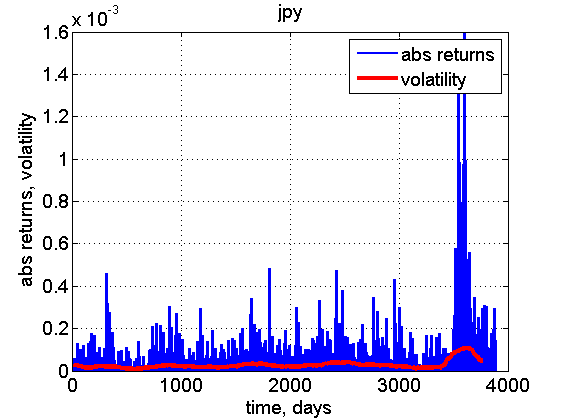}}
\caption{Absolute values of immediate speeds (abs returns) and their dispersions (volatility).}
\label{fig:Fig04}
\end{figure}

\begin{figure}[H]
\centerline{\includegraphics[width=6.16in,height=4.35in]{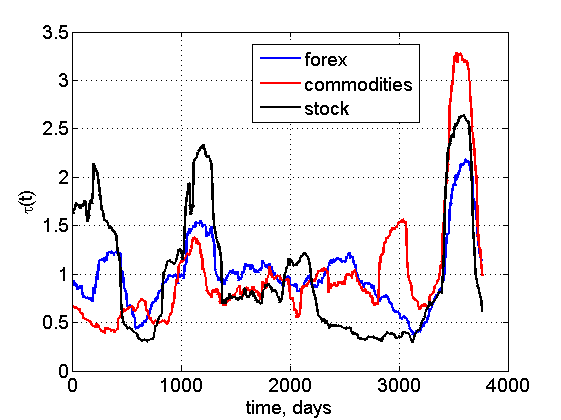}}
\caption{Coefficients of  time  compression-expansion, market ``temperature''. The explanation is in the text.}
\label{fig:Fig05}
\end{figure}

 Fig. \ref{fig:Fig05} shows averaged coefficients of time $\tau (t)$ compression-expansion  (formula (\ref{eq:GrindEQ__23_})) for three groups of incoming series: currency (forex), stock, and commodity markets. 

 The formulae (\ref{eq:GrindEQ__11_}),(\ref{eq:GrindEQ__23_}),(\ref{eq:GrindEQ__25_}) show that $\tau $(\textit{t}) exists in  proportion to the averaged square speed (according to the chosen time span and series), i.e. average ``energy'' of the economical ``particle'' (as it is in our analogy), and can be thus interpreted as the series ``temperature''. Crises are distinguished with the intensification of economic processes (the ``temperature'' is rising), while during the crisis-free period their deceleration can be observed (the ``temperature'' is falling), what can be interpreted as the heterogenous flow of economic time. $\tau (t)$ dependences shown on the fig. 5 illustrate all afore-mentioned. Note that local time acceleration-deceleration can be rather significant.

 Transition to heterogenous economic time allows to make the observed economic series more homogenous, which can simplify both analysis and prediction \cite{r042}.

 In table we give the values of a non-dimentional economic mass of the $m_{i} $ series, calculated using (\ref{eq:GrindEQ__22_}) for all 9 incoming series, as well as average masses of each group (formula (\ref{eq:GrindEQ__29_})).

 Table. Economic series masses

\begin{tabular}{|p{1.2in}|p{1.1in}|p{1.0in}|p{1.2in}|} \hline 
\multicolumn{2}{|p{1in}|}{Incoming series} & Economic mass & Average economic mass of the group \\ \hline 
\multirow{3}{*}{Commodity market} & gold & $2,816\cdot 10^4$  & \multirow{3}{*}{$4,983\cdot 10^3$} \\ 
 & silver & $4,843\cdot 10^3$  &  \\ 
 & oil & $2,777\cdot 10^3$  &  \\ \hline 
\multirow{3}{*}{Currency market} & jpy & $2,148\cdot 10^4$  &  \multirow{3}{*}{$2,499\cdot 10^4$}  \\ 
 & gbp & $3,523\cdot 10^4$  &  \\ 
 & chf & $2,180\cdot 10^4$  &  \\ \hline 
\multirow{3}{*}{Stock market} & S\&P 500 & $6,251\cdot 10^3$  & \multirow{3}{*}{$4,748\cdot 10^3$ } \\ 
 & FTSE 100 & $6,487\cdot 10^3$  &  \\ 
 & BVSP & $1,507\cdot 10^3$ &  \\ \hline 
\end{tabular}

 As you can see from the table, the stock market is distinguished with the lowest mass value, while the currency one shows the maximum number. Oil price series has the lowest mass on the commodity market, gold -- the highest one. As for the currency market, British pound (gbp) have the highest value and Japanese yen rates (jpy) demonstrates the minimum mass of the group, although the dispersion is lower than that of the commodity market. The smallest spread is peculiar to the currency market. Dynamic and developing Brazilian market (BVSP) has the lowest mass, while the maximum value, just like in the previous case, corresponds to Great Britain (FTSE 100). It is explained by the well-known fact: Britain has been always known for its relatively ``closed'' economy as comraped with the rest of the European and non-European countries. 

The last group of experimental data corresponds to the dependence of Planck's economic constant (calculated for different series) on time $\Delta t=k\Delta t_{\min } $ (time between the neighbouring registered observations), which is characterised by $H_{i} \left(k,\bar{n}\right)$ coefficient (see formula (\ref{eq:GrindEQ__24_})).

On fig. 6-8 integral dependencies  $H_{i} \left(k\right)$ are depicted. The following are averaged on the whole period of time 1993-2010 and calculated for commodity, stock, and currency markets. As you can see there are no obvious regularities, which can be explained by various crises and recessions of the world and national economies that took place during the investigated period. 

 To decide whether it is possible for local regularities of Planck's economic constant dependence on $\Delta t$ to appear, we have chosen relatively small averaging fragments, $\Delta N=500$, which approximately equals two years. Corresponding results for some of these fragments on commodity, currency, and stock markets are given on fig. 9-11. Evidently, all three figures show clear tendencies of $H_{i} \left(k,\bar{n}\right)$ recession and rise  for each type of the market (unlike integral dependences $H_{i} \left(k\right)$).

\begin{figure}[H]
\centerline{\includegraphics[width=6.16in,height=4.35in]{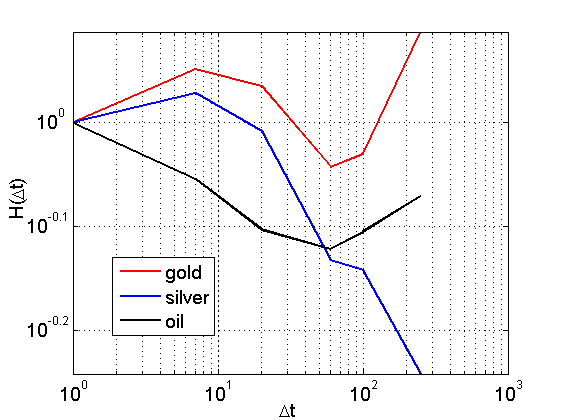}}
\caption{Integral coefficient $H_{i} \left(k\right)$ dependences for commodity market.}
\label{fig:Fig06}
\end{figure}

\begin{figure}[H]
\centerline{\includegraphics[width=6.16in,height=4.35in]{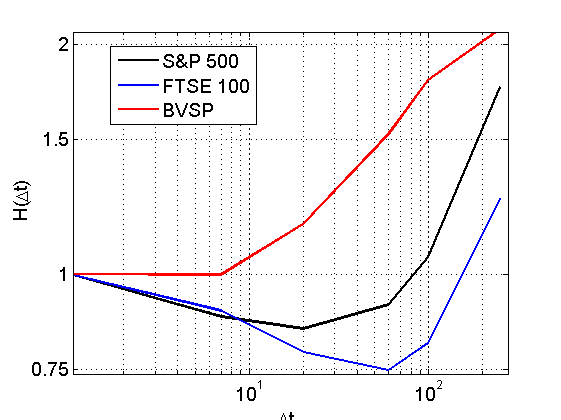}}
\caption{Integral coefficient $H_{i} \left(k\right)$ dependences for stock market.}
\label{fig:Fig07}
\end{figure}

\begin{figure}[H]
\centerline{\includegraphics[width=6.16in,height=4.35in]{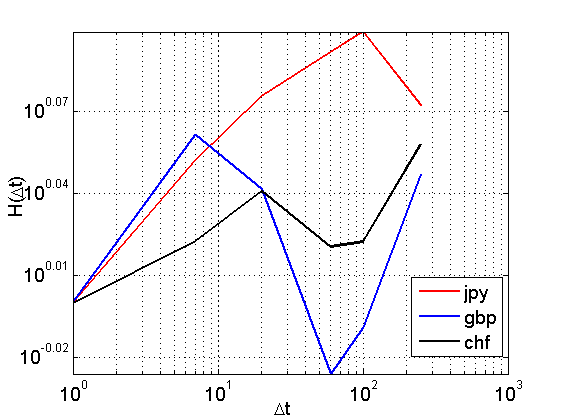}}
\caption{Integral coefficient $H_{i} \left(k\right)$ dependences for currency market.}
\label{fig:Fig08}
\end{figure}

\begin{figure}[H]
\centerline{\includegraphics[width=6.16in,height=4.35in]{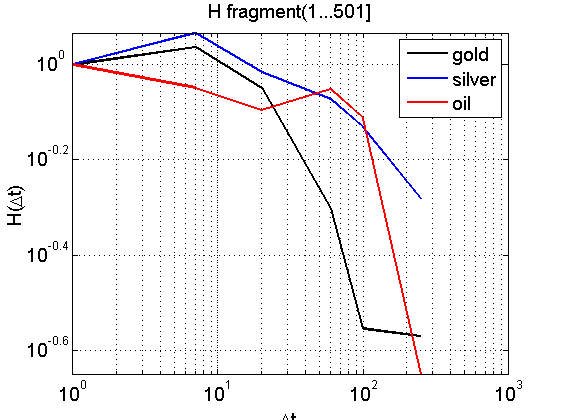}}
\caption{Local coefficient $H_{i} \left(k,\bar{n}\right)$ dependeces for commodity market  (averaging time span from 27.04.1993 to 12.06.1995, 500 daily values).}
\label{fig:Fig09}
\end{figure}

\begin{figure}[H]
\centerline{\includegraphics[width=6.16in,height=4.35in]{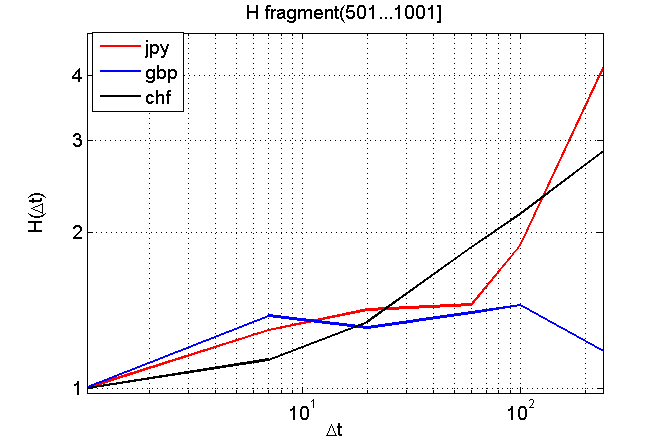}}
\caption{Local coefficient $H_{i} \left(k,\bar{n}\right)$ for currency market (averaging time span from 12.06.1995 to 15.07.1997, 500 daily values).}
\label{fig:Fig10}
\end{figure}

\begin{figure}[H]
\centerline{\includegraphics[width=6.16in,height=4.35in]{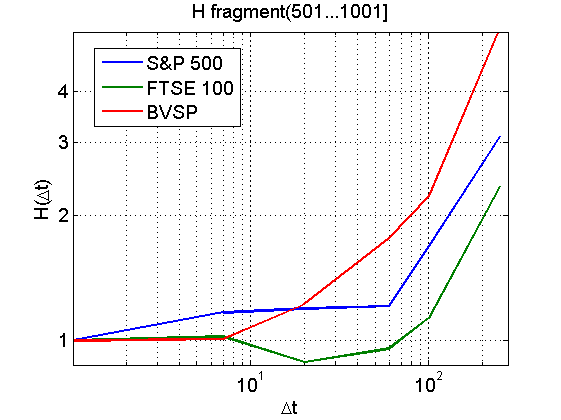}}
\caption{Local coefficient $H_{i} \left(k,\bar{n}\right)$ dependences for stock market (averaging time span from 12.06.1995 to 15.07.1997, 500 daily values).}
\label{fig:Fig11}
\end{figure}

\section{Conclusions}
\label{sec:Experiments}

 We have conducted methodological and philosophical analysis of physical notions and their formal and informal connections with real economic measurements. Basic ideas of the general relativity theory and relativistic qantum mechanics concerning spacetime properties and physical dimensions peculiarities were used as well. We have suggested procedures of detecting normalized economic coordinates, economic mass and heterogenous economic time. The afore-mentioned procedures are based on socio-economic time series analysis and economical interpretation of Heisenberg's uncertainty principle. The notion of economic Planck's constant has also been introduced. The theory has been tested on real economic time series, including stock indices, currency rates, and commodity prices. Acquired results indicate the availability of further investigations.

\bibliographystyle{unsrt}
\bibliography{Neopredelennost_arxiv_fromrus}

\end{document}